\documentclass[twocolumn,           
               showpacs,            
               preprintnumbers,     
               aps,                 
               prd,                 
               superscriptaddress,      
               nofootinbib,         
               tightenlines,        
               floats,floatfix      
               ]{revtex4}

\usepackage{graphicx}
\usepackage{color}
\usepackage{latexsym}
\usepackage{amsmath,amssymb}        
\usepackage[draft=false]{hyperref}

\begin{document}


\title{Interference pattern in the collision of structures in the BEC dark matter model: comparison with fluids}

\author{J. A Gonz\'alez}
\affiliation{Instituto de F\'{\i}sica y Matem\'{a}ticas,
        Universidad Michoacana de San Nicol\'as de Hidalgo. Edificio C-3,
        Cd. Universitaria,
        C. P. 58040 Morelia, Michoac\'{a}n, M\'exico.}

\author{F. S. Guzm\'an}
\affiliation{Instituto de F\'{\i}sica y Matem\'{a}ticas,
        Universidad Michoacana de San Nicol\'as de Hidalgo. Edificio C-3,
        Cd. Universitaria,
        C. P. 58040 Morelia, Michoac\'{a}n, M\'exico.}


\date{\today}


\begin{abstract}
In order to explore nonlinear effects on the distribution of matter 
during collisions within the Bose-Einstein condensate (BEC) dark matter 
model driven by the Schr\"odinger-Poisson system of equations, we study 
the head-on collision of structures and focus on the interference pattern 
formation in the density of matter during the collision process. We explore 
the possibility that the collision of two structures of fluid matter modeled
 with an ideal gas equation of state also forms interference patterns and 
found a negative result. Given that a fluid is the most common flavor of 
dark matter models, we conclude that one fingerprint of the BEC dark matter 
model is the pattern formation in the density during a collision of 
structures.
\end{abstract}


\pacs{95.35.+d, 05.30.Jp, 03.75.Lm, 05.45.Yv}


\maketitle



\section{Introduction}

The discovery of the nature of dark matter is one of the most important 
problems nowadays, and a considerable amount of theoretical models have
been proposed. One of the candidates that has received certain attention 
is a scalar field dark matter ultralight particle that is consistent with 
cosmological observations \cite{Varun,MatosUrena2000}. At cosmological 
scales, this and other related  models have been widely studied, and one of 
the important issues to be understood about this candidate is its behavior 
at local scales related to nonlinear processes of collapse, structure 
formation and collision of structures. Some theoretical studies have 
pushed forward this model at local scales, especially those related to  
galactic rotation curves under various conditions \cite{Lee,RCs,Arbey}. 
In the case of ultralight scalar fields, it has been shown that the density 
profiles of scalar field structures is not cuspy as opposed to Navarro 
Frenk White density profiles \cite{BernalMatosNunez}, and also predicts 
an adequate cutoff in the mass power spectrum of structures consistent 
with the abundance of small structures \cite{MatosUrena2000}. Finally, more recently the evolution of cosmological perturbations in Bose-Einstein condensate dark matter is already under study \cite{Harko2011}.

In the nonlinear regime, some advances have been made related to the 
evolution of gravitating scalar field structures, both relativistic under 
general relativistic conditions \cite{Oscillatons} and Newtonian in the 
low energy and weak gravitational field regime \cite{BECspherical,
BECspherical2,BernalGuzman2006a,BernalGuzman2006b}.

The Newtonian case has been found to be more appropriate to study the 
nonlinear collapse and interaction between structures, and is driven 
by the time-dependent Schr\"odinger-Poisson (SP) system of equations. 
The interpretation of this system of equations is that Schr\"odinger 
equation represents a Bose-Einstein condensate (BEC) of the scalar 
field particles at zero temperature in the mean field approximation, 
through the Gross-Pitaevskii equation \cite{GrossPitaevskii}. This is 
the reason why in the low energy and weak gravitational field limits, 
the scalar field dark matter model is called BEC dark matter.

The solution of the time-dependent SP system requires the application 
of numerical methods to visualize the evolution of BEC structures and 
estimate any potential observable implication. A helpful feature of the 
SP system is that it has equilibrium configurations, that is, there are 
spherically symmetric stationary solutions based on the assumption that 
the wave function depends harmonically in time, which in turn implies 
that both the gravitational potential and the energy density are 
time-independent \cite{BECspherical,BECspherical2}.
Such equilibrium configurations have been studied dynamically and found 
to be stable under spherical  perturbations and are also attractor 
solutions in time for initial perturbations with arbitrary profile  
\cite{BECspherical2}; they are also stable against nonspherical 
perturbations \cite{BernalGuzman2006a}, and they serve as stable structures 
to study how BEC dark matter structures behave in nonlinear situations.

In this paper, we explore the BEC dark matter model during the head-on 
collision of structures as depicted first in \cite{BernalGuzman2006b}, 
where it was found that when the binary system is initially bounded, the 
resulting system is a single object, whereas for unbounded systems a sort 
of solitonic behavior was shown to happen. In both cases, it was found that 
an interference pattern was formed during the collision which potentially 
would provide predictions on the behavior of baryonic matter whose stream 
would be driven in part by the dark matter gravitational potential during 
the collision of two structures.

On the other hand, the most studied models of dark matter assume it can 
be a dust fluid, which is an approximate model of dark matter candidates 
like WIMPs. Thus, in this work we present a comparison of the head-on 
collision of two structures assuming on the one hand that the structures 
correspond to two equilibrium configurations of the Schr\"odinger-Poisson 
system and on the other hand the structures correspond to two relaxed balls 
of ideal gas. 

The goal is to determine whether or not the collision of two balls of gas 
representing two structures of WIMPs can also present an interference 
pattern during the head-on collision, and thus conclude that the BEC dark 
matter has peculiar fingerprint interference patterns during the collision 
of structures that should predict observations that determine its viability,
or determine the BEC dark matter should be ruled out.

This paper is organized as follows: In Sec. II we introduce the systems 
of equations we solve, and the numerical methods we use in order to carry 
out the head-on collision of two structures, both of BEC dark matter and 
of an ideal gas; in Sec. III we present the results of our simulations 
and compare the behavior in the two cases; and finally in Sec. IV we draw 
some conclusions.


\section{Systems of equations and numerical methods}

\subsection{The Schr\"odinger-Poisson system}

{\it The system of equations.} The SP system of equations consists of 
the Schr\"odinger equation for a wave function $\psi$, with a potential 
that is solution of Poisson equation sourced by the density of probability 
$|\psi|^2$. Since we deal with a head-on collision, we write down the SP 
system in cylindrical coordinates as follows:

\begin{eqnarray}
i \frac{\partial \psi}{\partial t} &=& -\frac{1}{2} \left(
                                \frac{\partial^{2} \psi}{\partial x^2}
                              + \frac{1}{x}\frac{\partial \psi}{\partial x}
                              + \frac{\partial^{2} \psi}{\partial z^2}
                                \right)
                    + U \psi + \Lambda|\psi|^2\psi\label{eq:schroedinger}\\
\frac{\partial^{2} U}{\partial x^2} &+&
\frac{1}{x}\frac{\partial U}{\partial x} +
\frac{\partial^{2} U}{\partial z^2} =
\psi^{\ast}\psi,\label{eq:poisson}
\end{eqnarray}

\noindent where $\psi=\psi(x,z,t)$ and $U=U(x,z,t)$ are the wave function 
and the gravitational potential respectively; $x,z$ are the radial and 
axial cylindrical coordinates respectively. The third order term in   
Eq.~(\ref{eq:schroedinger}) is related to a self-interacting term, in 
which $\Lambda$ corresponds to the s-wave scattering length in the 
Gross-Pitaevskii approximation for Bose condensates \cite{GrossPitaevskii}, 
which we set to zero in this work, because it is irrelevant in the 
interference pattern formation studied here. This term instead was shown 
to play the role of determining the compactness of an equilibrium 
configuration \cite{BECspherical2}. Eqs. (\ref{eq:schroedinger}-
\ref{eq:poisson}) use the units and scaling $\hbar = c = 1$ with 
$x\rightarrow mx$, $z\rightarrow mz$, $t \rightarrow mt$ and the wave 
function $\psi \rightarrow \sqrt{4\pi G}\psi$, where $m$ is the mass of 
the ultralight boson. A consequence of this change of units is that 
the mass of a system will be in units of $[M]=M^{2}_{pl}/m$, and thus $m$ 
determines the scale of the configurations we start with. In fact these 
units define the Kaup mass, which in the relativistic counterpart of the 
Schr\"odinger-Poisson system (the boson star case ruled by the 
Einstein-Klein-Gordon system of equations) indicates the threshold between 
stable and unstable bounded configurations \cite{Kaup}. It is worth 
mentioning that in the case of the Schr\"odinger-Poisson system, that is, 
the low energy and weak field limits of boson stars, the equilibrium 
configurations are all stable, unless a negative self-interaction factor 
$\Lambda$ is introduced \cite{BECspherical2}.

{\it The evolution of the system.} We consider the system 
(\ref{eq:schroedinger}-\ref{eq:poisson}) to be a constrained evolution 
system, that is the Schr\"odinger equation is an evolution equation that 
satisfies the Poisson equation which is a constraint. Explicitly, we 
integrate the Schr\"odinger equation using a finite differences approximation 
along the spatial directions and a method of lines for the 
integration in time that uses a third order Runge-Kutta algorithm. We 
solve the constraint (Poisson equation) at the intermediate steps of 
the time integrator, which provides a full coupling of the evolution 
constrained system, as shown in our convergence tests. The flow of the 
solution is as follows: at every intermediate iteration of the Runge-Kutta 
integrator: i) evolve the system using the Schr\"odinger equation to evolve 
$\psi$, ii) use such updated value of the wave function to source and 
solve the Poisson equation, iii) then obtain a new potential $U$ and repeat. 
Usually the Schr\"odinger equation is integrated in time using unitary 
operators \cite{Choi}, which is related to fully implicit methods, 
however, even though we use explicit methods we verified the evolution 
is unitary as shown in \cite{BECspherical2,BernalGuzman2006a,
BernalGuzman2006b}. In order to avoid the singularity at $x=0$ in 
Eqs. (\ref{eq:schroedinger}-\ref{eq:poisson}) with our finite 
differences approximation, we stagger the axis such that we avoid the 
origin, and in order to achieve the second order convergence at the axis
we use the identity $\frac{1}{x}\frac{\partial f}{\partial x}=
2\frac{df}{dx^2}$, and code the later expression, which is a derivative 
with respect to $x^2$.

{\it Poisson equation.} Eq. (\ref{eq:poisson}) is an elliptic 
equation for $U$ which we solve using the 2D five-point stencil for 
the derivatives and a successive over-relaxation iterative 
algorithm with optimal acceleration parameter \cite{Smith1965}. In order 
to impose boundary conditions we made sure the boundaries are far enough 
for the number of particles represented by the integral of the density 
of probability $N=\int |\psi|^2 d^3x$ to be the same along the boundary 
of the domain and used a monopolar term of the gravitational field; 
that is, we used the value $U = -N/r$ along the boundaries with 
$r=\sqrt{x^2+z^2}$. At the axis we demand the gravitational potential 
to be symmetric with respect to the axis.

{\it Boundary conditions during the evolution.} We use a sponge in the 
outermost region of the domain. The sponge is a concept used with success 
in the past when dealing with the Schr\"odinger equation \cite{Israeli1981,
GuzmanUrena2004}. This technique consists in adding up to the potential 
in the Schr\"odinger equation an imaginary part. The result is that in 
the region where this is applied there is a sink of particles, and 
therefore the density of probability in this region will be damped out, 
with which we get the effects of an outgoing flux boundary condition 
\cite{GuzmanUrena2004}. The expression we use for the sponge profile is

\begin{equation}
V = -\frac{i}{2} V_0 \left\{ 2 + \tanh \left[(r_{jk}-r_c)/\delta
\right] - \tanh \left( r_c/\delta \right) \right\} \, , \label{imagpot}
\end{equation}

\noindent which is a smooth version of a step function with amplitude 
$V_0$, centered at $r_c$ and width $\delta$; $r_{jk}=
\sqrt{x_{jk}^{2}+z_{jk}^{2}}$ is the radius of a given point of the 
discretized domain; the minus sign guarantees the decay of the number 
of particles at the outer parts of our integration domain, that is, 
the imaginary potential behaves as a sink of particles.

{\it Initial data.} In order to reduce our parameter space to specific 
but illustrative cases, we only consider the equal mass head-on collision 
case. Thus we construct the solution of a spherically symmetric 
equilibrium ground state configuration in spherical coordinates as done 
in \cite{BECspherical,BECspherical2,GuzmanUrena2004}. In the units used 
the mass of each configuration is $N=2.06$ with a radius containing the
95\% of the total mass is $x_{95}=3.93$ \cite{GuzmanUrena2004}. We  obtain 
a wave function $\psi_{sph}$ and gravitational potential $U_{sph}$. We then 
interpolate such wave function at two different places of the two-dimensional 
grid $(0,\pm z_0)$ centered along the $z$-axis and define a global wave 
function $\psi=\psi_1+\psi_2$, where $\psi_1$ and $\psi_2$ are the wave 
functions obtained from the spherical solution located at two different 
places of the 2D grid. We choose $z_0$ such that the two configurations 
are far enough one from the other so that the interference term 
$<\psi_1,\psi_2>$ lies near to round off error values, in order to 
consider the two blobs have the adequate phase; the magnitude of this 
term decreases exponentially with the initial separation of the blobs. 
In this way, we solve the Poisson equation sourced by the energy 
density $\rho=|\psi|^2$. Then we have initial data for two superposed 
ground state equilibrium configurations in our axially symmetric domain.

{\it Analysis.} In order to analyze the kinetic development of the 
collision it is important to track global quantities during the evolution, 
such as kinetic, gravitational and total energy. We do this by calculating 
the expectation values 

\begin{eqnarray}
K &=& -\frac{1}{2}\int \psi^{\ast} \nabla^2 \psi d^3x,\label{eq:expectationK}\\
W &=& \frac{1}{2}\int \psi^{\ast}U\psi d^3x,\label{eq:expectationW}
\label{eq:expectationI}
\end{eqnarray}

\noindent which are respectively the expectation values of the kinetic 
and gravitational energies. These quantities are important at 
determining the state of the system at any time during the evolution of 
the system. For instance, the value of the total energy $E = K+W$ 
indicates whether we account with a bounded system or not, and the 
very important virial theorem relation $2K+W=0$, which is nearly 
satisfied when the system gets virialized and relaxed through whatever 
channels available, for instance, the emission of scalar field bursts, 
the so-called gravitational cooling process \cite{Seidel}.

\subsection{The SPH algorithm for the fluid}

We want to study the head-on collision of two structures of ideal gas. 
In order to do that, we need to solve Euler's equations coupled to
Newtonian gravity. With this in mind, we implemented a smoothed 
particle hydrodynamic (SPH) code based on the formulation described 
in ~\cite{Rosswog}.
This scheme allows us to evolve the position, velocity, density, 
pressure and internal energy of the fluid elements. We explore two 
different scenarios of ideal. In the first one we set the internal 
energy and pressure equal to zero, and in the second one we use the 
more general ideal gas equation of state $p=(\Gamma -1)\rho u$, where 
$p$ is the pressure of the gas, $\rho$ is its density, $\Gamma$ is a 
constant and $u$ is the internal energy of the gas.

{\it Initial data.} The initial values required for our simulations 
are the positions, velocities, density, pressure and internal energy 
of the particles in the fluid.

We start choosing the density profile given by the Plummer model:

\begin{equation}
\rho(r) = \frac{3M}{4\pi R^3} \left( 1+\frac{r^2}{R^2}\right)^{-5/2} \, ,
\end{equation}

\noindent for the fluid without pressure and the density profile:

\begin{equation}
\rho(r) = \frac{M}{2\pi R^2} \frac{1}{r} \, ,
\end{equation}

\noindent for the fluid with pressure.

We construct one of the structures using an acceptance-rejection method 
to generate the position of the particles in such a way that they satisfy 
the corresponding density profile. We set the initial velocities of 
the particles equal to zero and set the parameters $M=2.0622$ and $R=3.93$. 
Finally, we set the value of the internal energy equal to a constant 
($u=0$ for the pressureless fluid and $u=0.05M/R$ in the other case) 
and recover the pressure from the equation of state. Such configurations 
are not in equilibrium, therefore we evolve them numerically and let 
them settle down into an equilibrium state.

In order to generate the initial data corresponding to the head-on 
collision of two bumps of fluid, we take the resulting equilibrium 
configuration described in the previous paragraph, make two copies 
of it and place them at a given separation along the $z$-axis. Then 
a boost on each of the configurations is applied along the head-on 
axis. The initial positions are $z = \pm 10.0$ and the initial 
velocity is $v_z^0 = \pm 2.6875/2.0622$. This value was chosen such 
that the total energy of the system is equal to zero, although other 
values of the boost were explored and the results are qualitatively 
the same.

\section{Results}

We perform a series of head-on collisions of two initial balls in 
both cases, the one driven by the SP system of equations and the 
fluid driven by hydrodynamics. We set up configuration for the SP 
system and for the fluid that are dynamically comparable in time, 
size and separation. The reason is that there is no way to identify 
exactly an initial configuration corresponding to the quantum 
mechanical system driven by the Schr\"odinger equation (which is 
dispersive) with the configurations constructed for an ideal gas.

We define the total energy of the system as $E=K+W$, where $K$ and 
$W$ are the expectation values of the kinetic gravitational energies 
defined in (\ref{eq:expectationK}-\ref{eq:expectationW}), calculated 
with the time-dependent wave function that is being calculated on the 
fly of the Schr\"odinger-Poisson system, and $E=K+U+W$ is the total 
energy, that is the sum of the kinetic, internal and gravitational 
energies in the case of the fluid. The total energy depends strongly 
on the value of the initial head-on momentum the initial balls have, 
so that we choose this parameter to be the one we tune in order to 
obtain our initial binary configurations.

The parameter space we explore corresponds to initial configurations 
with different values of the total energy $E$, which allows to 
explore situations in which the system is both bounded ($E<0$) and 
unbounded ($E>0$).

We consider there is a pattern formation if there is a sequence of 
strips of the density of matter of the size of the order of the 
initial radii of the initial configurations as shown to happen 
during the collision of two blobs of BEC \cite{BernalGuzman2006b}.

\subsection{Bose condensate initial configurations}

We choose two ground state spherically symmetric equilibrium 
configurations that are originally virialized, that is, they satisfy 
separately the condition $2K+W=0$ and superpose them into the 2d grid 
(for details in the construction of equilibrium configurations see 
\cite{GuzmanUrena2004}). We apply momentum along the head-on direction 
in the following way: assume  $\psi=\psi_1+\psi_2$ is the wave function 
resulting from the superposition of the two equilibrium configurations, 
where $\psi_1$ and $\psi_2$ represent the wave function of each of the 
two equilibrium configurations centered along the head-on axis $z$ at 
$\pm z_0$, being $\psi_1$ and $\psi_2$ centered at $-z_0$ and $z_0$ 
respectively. Thus we add up momentum to each of the configurations 
defining a new wave function $\psi=e^{ip_z}\psi_1 + e^{-ip_z}\psi_2$. 
After this process we solve Poisson equation and start the evolution.

According to the Gross-Pitaevskii theory, BECs behave like classical waves, 
and as such interference phenomena are expected. Experiments showing 
interference patterns are well known \cite{Andrews1997}. In these 
experiments two separate Bose condensate configurations are released 
from their respective traps and allow them to evolve freely; 
eventually the two densities of probability interfere and produce a 
total density of the form

\begin{equation}
|\psi_1 + \psi_2|^2 = |\psi_1|^2 + |\psi_2|^2 + 2 <\psi_1,\psi_2>\cos
\left( \frac{md}{\hbar t} + \Phi \right),
\label{eq:experiments}
\end{equation}

\noindent where $m$ is the mass parameter in Schr\"odinger's equation, 
and $d$ is the initial distance between the two condensates, and $\Phi$ 
is the initial phase difference between the two condensated given by 
$\psi=\psi_1 + e^{i\Phi}\psi_2$, which is nonzero when there is a 
nonzero head-on initial momentum. The density of probability
(\ref{eq:experiments}) predicts a spacing between two consecutive 
fringes of constructive interference given by

\begin{equation}
\lambda = \frac{ht}{md},
\label{eq:lambda}
\end{equation}

\noindent which in our units corresponds to $\lambda = 2\pi t /d = 
2\pi t /(2 z_0)$.  This type of pattern formation has also been 
verified with simulations of laboratory systems (see e.g. 
\cite{Rohrl1997}). We have to remind the reader that an important 
difference between the conditions in these experiments and our 
system is that we actually never release the condensates from their 
respective traps which are the gravitational potentials the density of probability generates, instead, we allow the traps to interact with each 
other. Nevertheless we use the formula to measure the distance 
between fringes and find that for big values of the initial head-on 
momentum $p_z$, that is, a big value of $\Phi$ in the experiments, 
the law is nearly valid, whereas for small phases it does not hold. 
We interpret this result as a consequence of the fact that for the cases where the head-on momentum is small, a merger is actually 
expected to occur and the interaction between the traps becomes 
important, whereas in the high momentum a nearly solitonic 
trespassing effect happens. 

\begin{figure}[htp]
\includegraphics[width=4cm]{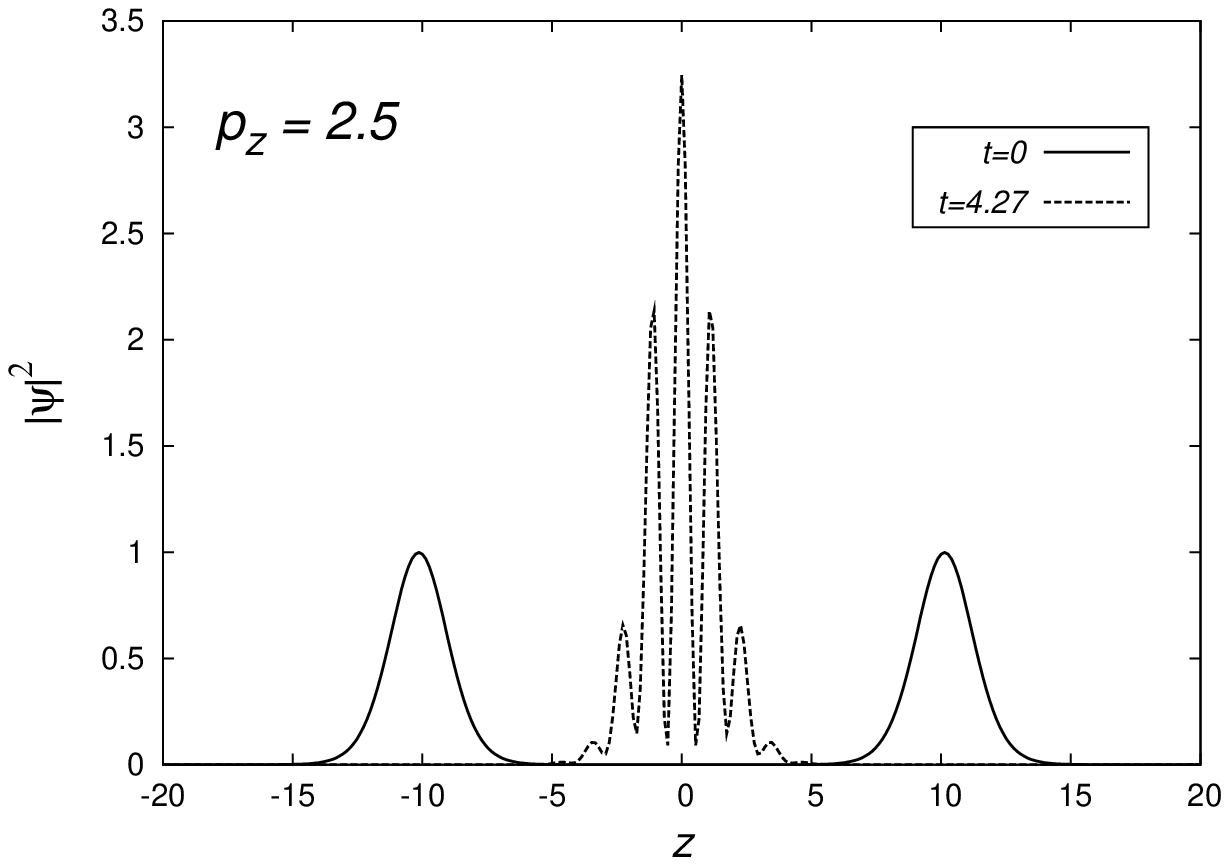}
\includegraphics[width=4cm]{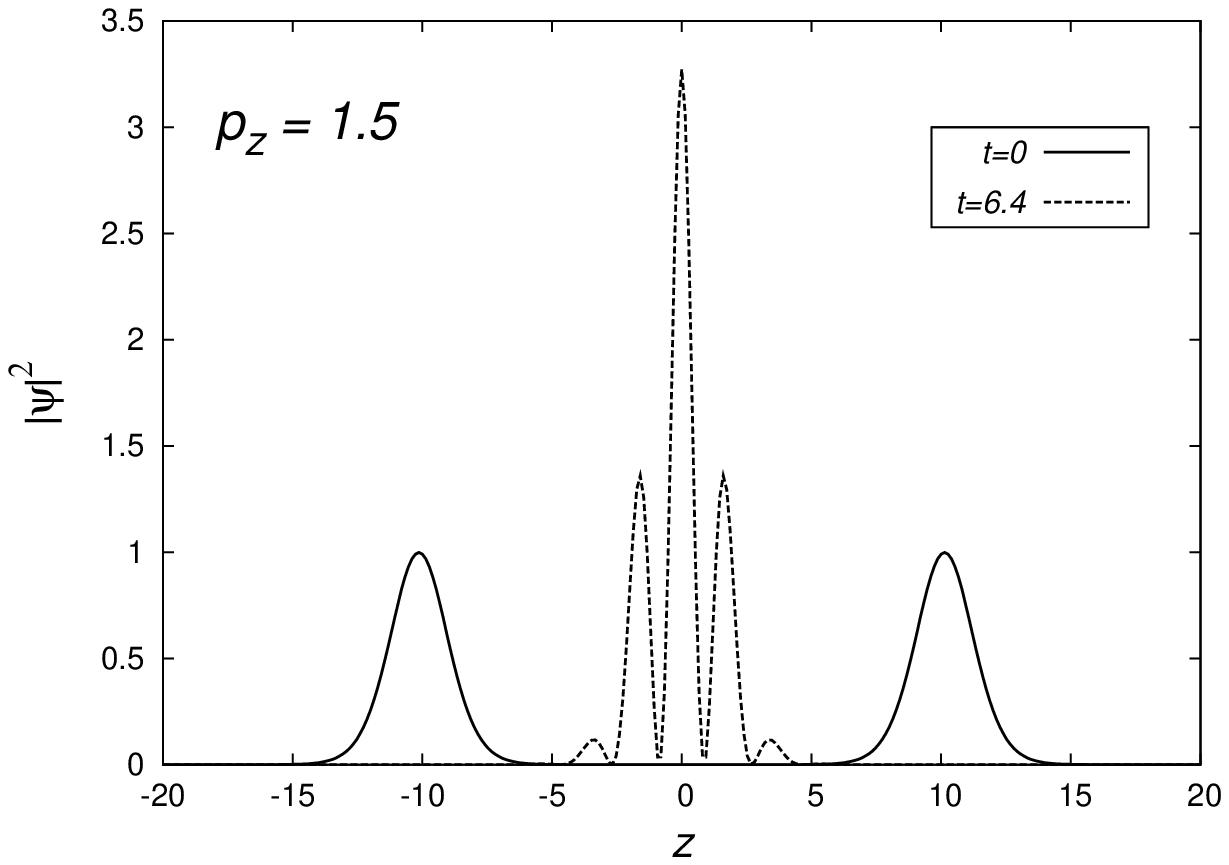}
\includegraphics[width=4cm]{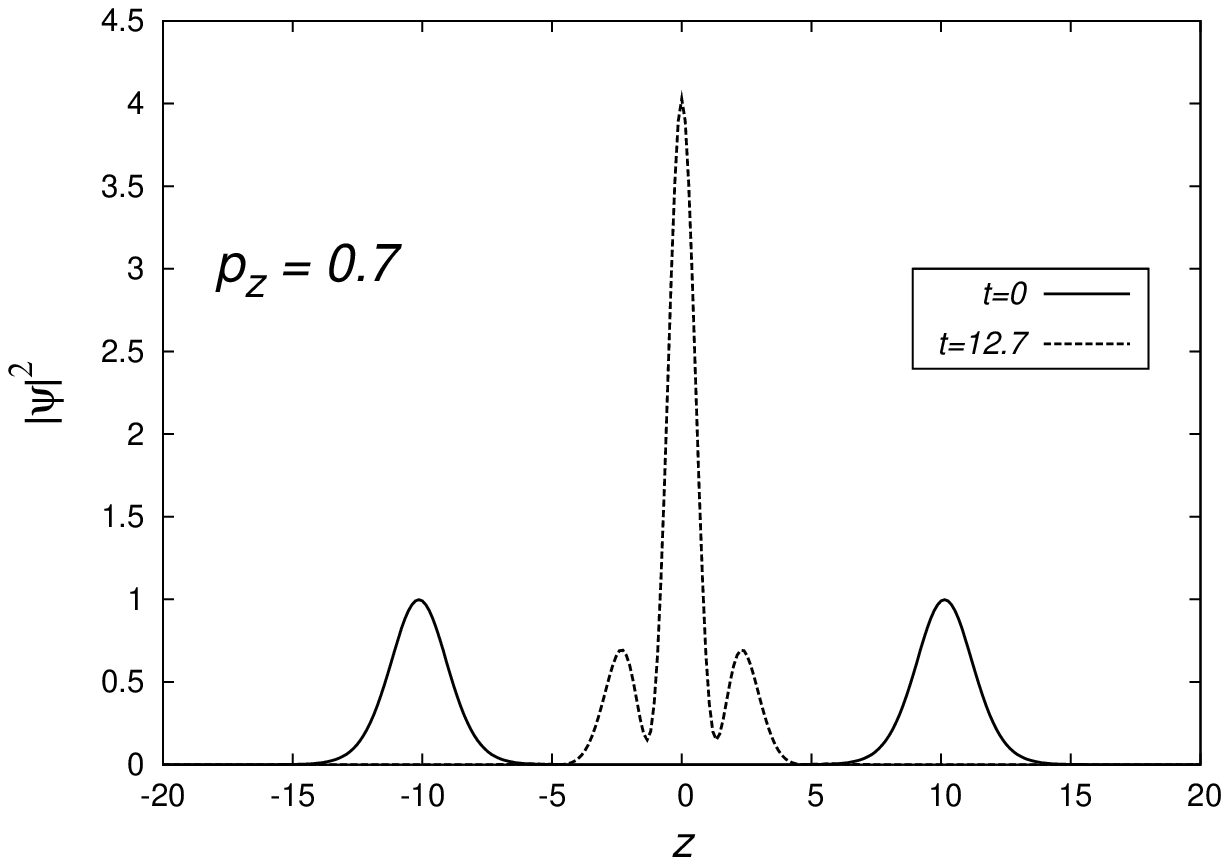}
\includegraphics[width=4cm]{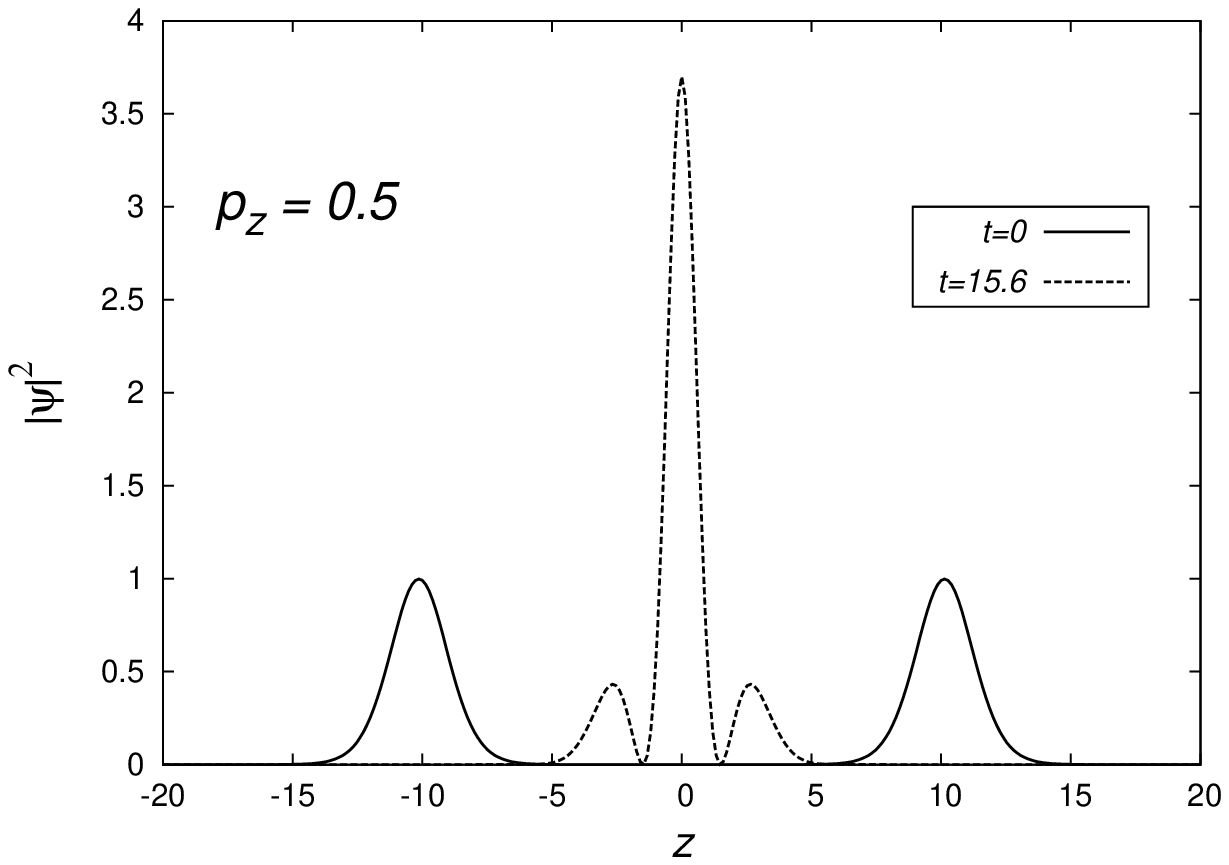}
\includegraphics[width=4cm]{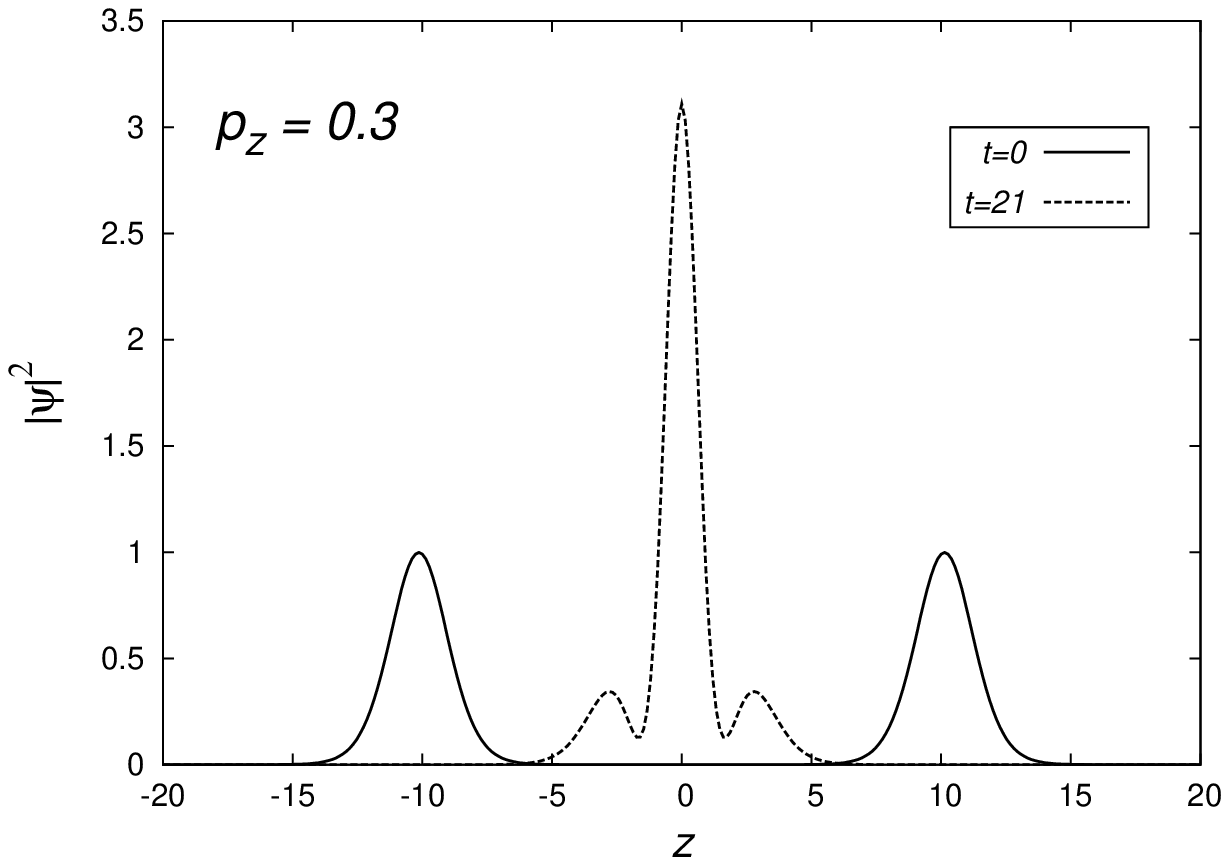}
\includegraphics[width=4cm]{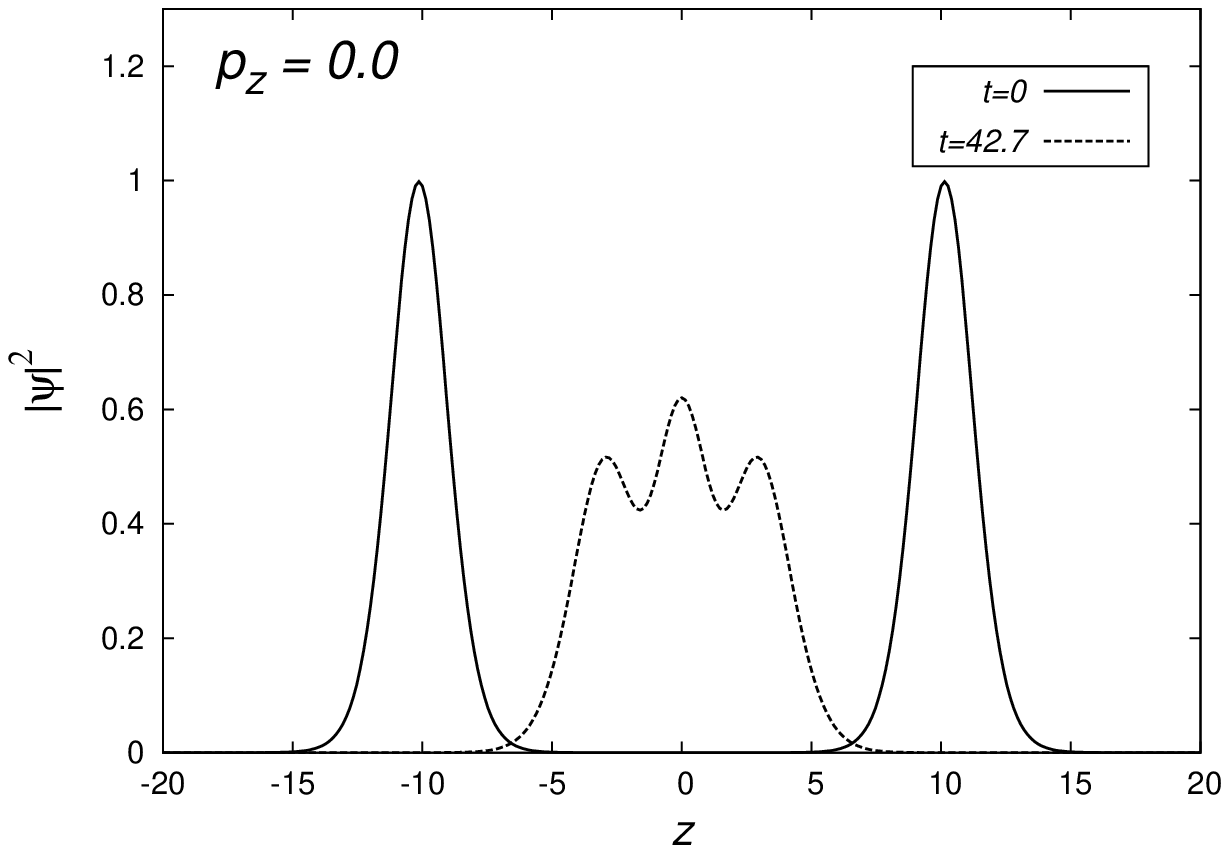}
\caption{\label{fig:bec_patterns} We show the best snapshot of the 
density $|\psi|^2$ along the $z$ axis showing the interference pattern. 
We also show the initial value of the density, in order to compare 
the size of the interference pattern.}
\end{figure}

The parameter space we consider is shown in Table \ref{table:bec} 
and snapshots of pattern formation are shown in Fig. 
\ref{fig:bec_patterns}. Notice that the width of the energy density 
in the patterns is even thicker than the initial size of the 
equilibrium configurations at initial time.

\begin{table}[htb]
\begin{tabular}{|c|c|c|c|c|}
\hline
$p_z$	& sgn(E)	& time for collision	& $\lambda$ given by 
(\ref{eq:lambda}) & $\lambda$ (measured) \\\hline
0.0		& -		& $\sim 43$		& 13.5	& 2.9\\
0.3		& -		& $\sim 21$		& 6.59	& 2.8\\
0.5		& -		& $\sim 16$		& 5.03	& 2.7\\
0.7		& 0		& $\sim 13$		& 3.99	& 2.3\\
1.5		& +		& $\sim 6.5$		& 2.01	& 1.8\\
2.5		& +		& $\sim 4.3$		& 1.27	& 1.2 \\\hline
\end{tabular}
\caption{ \label{table:bec} Parameters of the initial configurations 
for the BEC case. We also show the value of the distance between 
fringes of interference both, as predicted by formula (\ref{eq:lambda}) 
for laboratory experiments and the one we measure.}
\end{table}

\begin{figure}[htp]
\includegraphics[width=7.5cm]{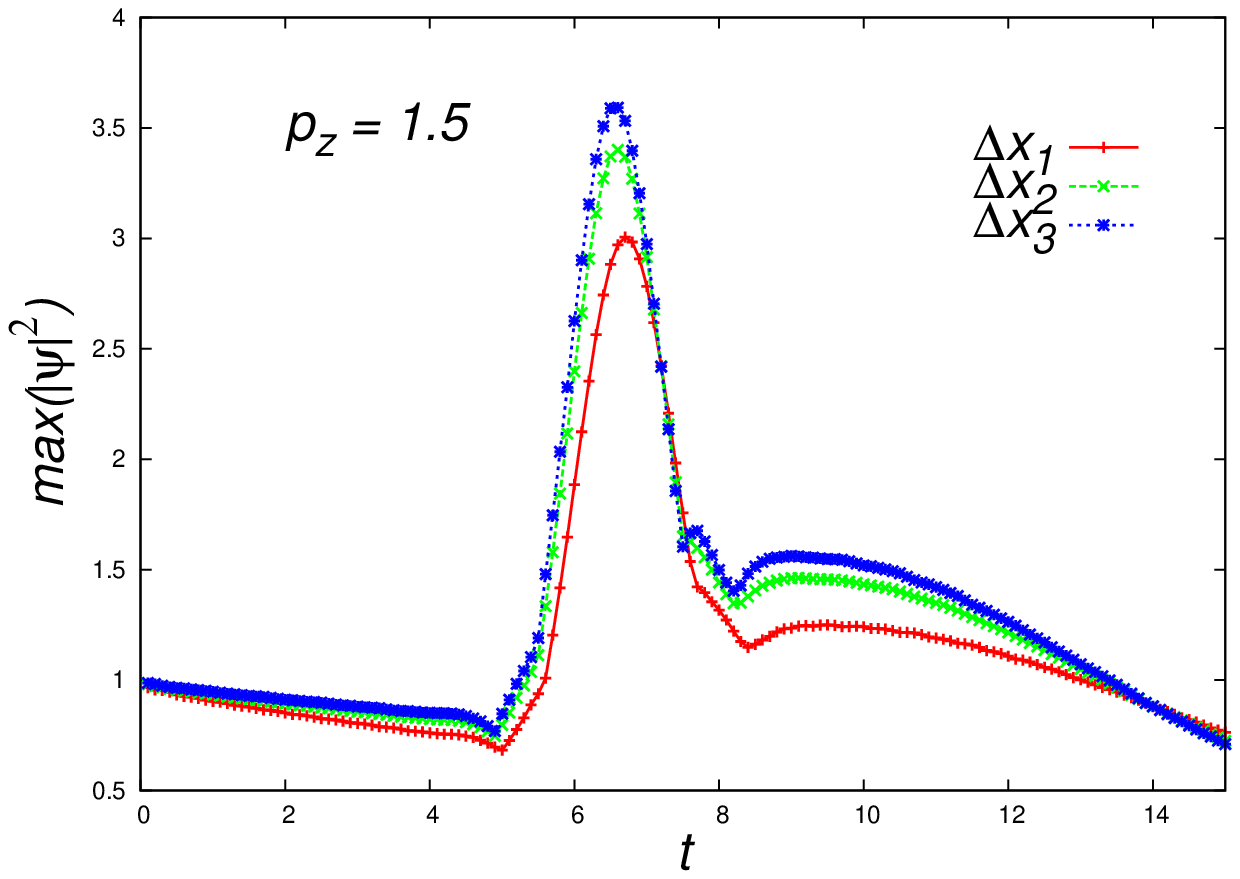}
\includegraphics[width=7.5cm]{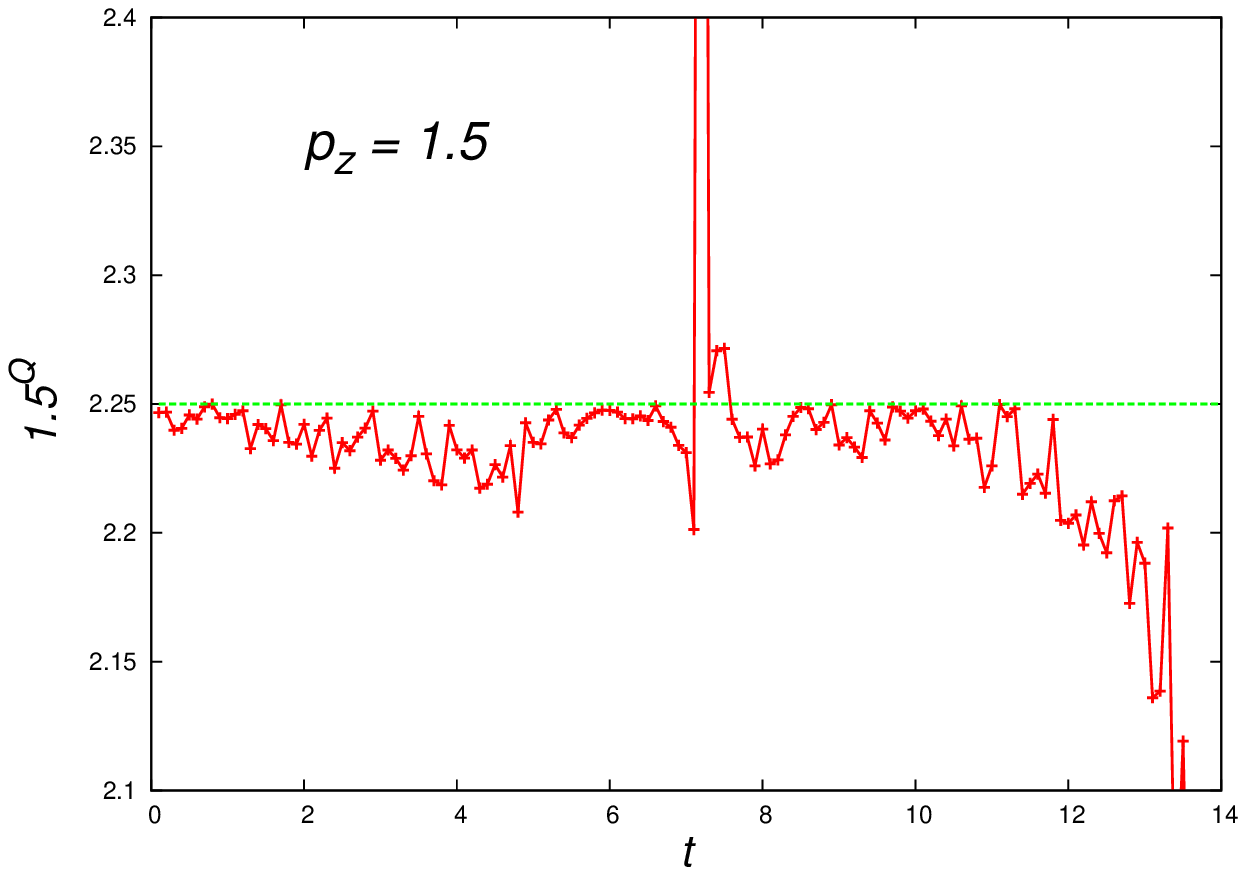}
\caption{\label{fig:bec_convergence} (Top) In order to show the 
convergence of our algorithms we track the maximum of the energy 
density during the evolution for the particular case $p_z=1.5$, which 
would be equivalent to estimate the infinity norm of the energy density. 
In our calculations we used three different resolutions $\Delta x_1=0.2$, 
$\Delta x_2 = \Delta x_1/1.5$ and $\Delta x_3= \Delta x_2/1.5$. (Bottom) 
We show the convergence factor, that is $1.5^Q$, which in theory should 
be $1.5^2=2.25$ for a second order convergent implementation. We show 
how much our calculations approach second order convergence; the 
convergence factor oscillates usually near or around the theoretical 
value due to phases in the scalar (the maximum of the density in our 
case) for the various resolutions. A big peak also appears approximately 
by the time the density approaches its maximum, and we see how convergence 
is lost after around $t \sim 11$ and decreases clearly afterwards by the 
time the blobs reach the sponge region, where the calculations are not 
expected to converge since there Schr\"odinger's equation has been 
modified with an artificial sink of particles, and also the unitarity 
of the evolution in the numerical domain is lost.}
\end{figure}

We perform our calculations using the domain $x \in [0,20]$, $z \in 
[-20,20]$ and a uniformly discretized grid, with constant time 
resolution $\Delta t$ and the same resolution along both spatial 
directions $\Delta x = \Delta z = \Delta xz$ and a sponge radius 
with $r_{c}=17$ and $\delta=1$. In order to validate our numerical 
results we show in Fig. \ref{fig:bec_convergence} a criterion that 
shows our results converge. As mentioned above, we use a method of 
lines with second order stencils along the spatial directions and a 
third order accurate integrator in time, so that we expect the code 
to converge to second order at least. Since we have no exact solution 
to our problem to compare with, or a constraint of the PDE system of 
equations we should expect to be satisfied, we can only practice a 
self-convergence test, that is, we use three different resolutions 
$\Delta xz_1=0.2$, $\Delta xz_2 =0.13333= \Delta xz_1/1.5$ and 
$\Delta xz_3=0.08888= \Delta xz_2/1.5$ and track the maximum of the 
energy density $\rho_{bec}=|\psi|^2$, which corresponds to the infinity 
norm of $\rho_{bec}$ because it is always non-negative. Given the ratio 
between our resolutions is $1.5$, we expect the maximum of the density 
to satisfy 

\begin{equation}
\frac{max(\rho_{bec}[using ~ \Delta x_1]) - max(\rho_{bec}[using ~ 
\Delta x_2])}{max(\rho_{bec}[using ~ \Delta x_2]) - max(\rho_{bec}[using ~ 
\Delta x_3])} = 1.5^Q,
\end{equation}

\noindent where $Q$ is the order of convergence expected to hold, 
which is at least two in our case because such is the order of 
accuracy of our stencils along the spatial directions and at most 
three, which is the accuracy of our time integrator. In Fig. 
\ref{fig:bec_convergence} we also show the value of $Q$. In all our 
calculations we kept the factor $\Delta t /\Delta xz^2 =0.1$ constant 
for all the resolutions, which maintains the evolution stable, 
accurate and convergent.

\subsection{Fluid configurations}

We evolve the initial configurations of fluid integrating Euler's equations 
using a predictor-corrector scheme.
We compute the total, kinetic, potential and internal energy of the system
in order to monitor the behavior of our numerical evolution.

\begin{figure}[htp]
\includegraphics[width=4cm]{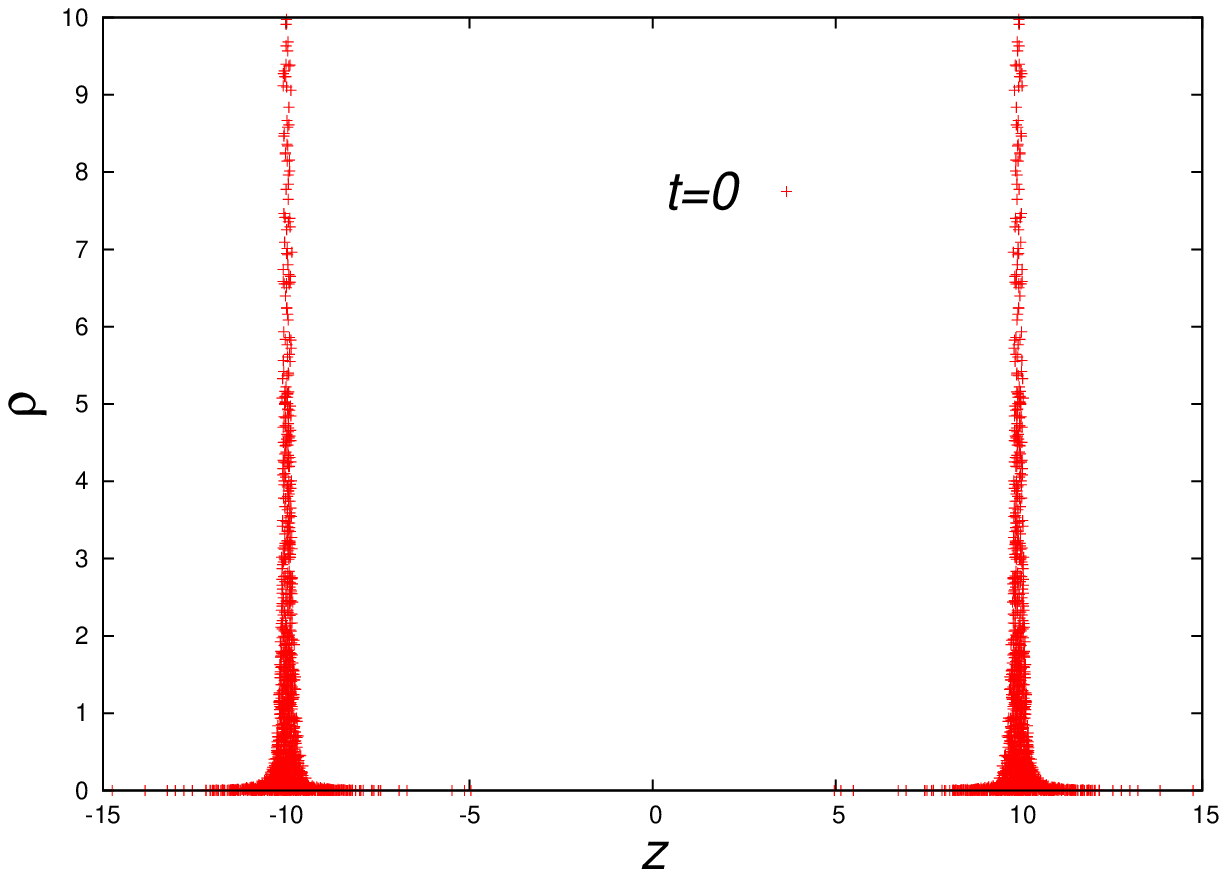}
\includegraphics[width=4cm]{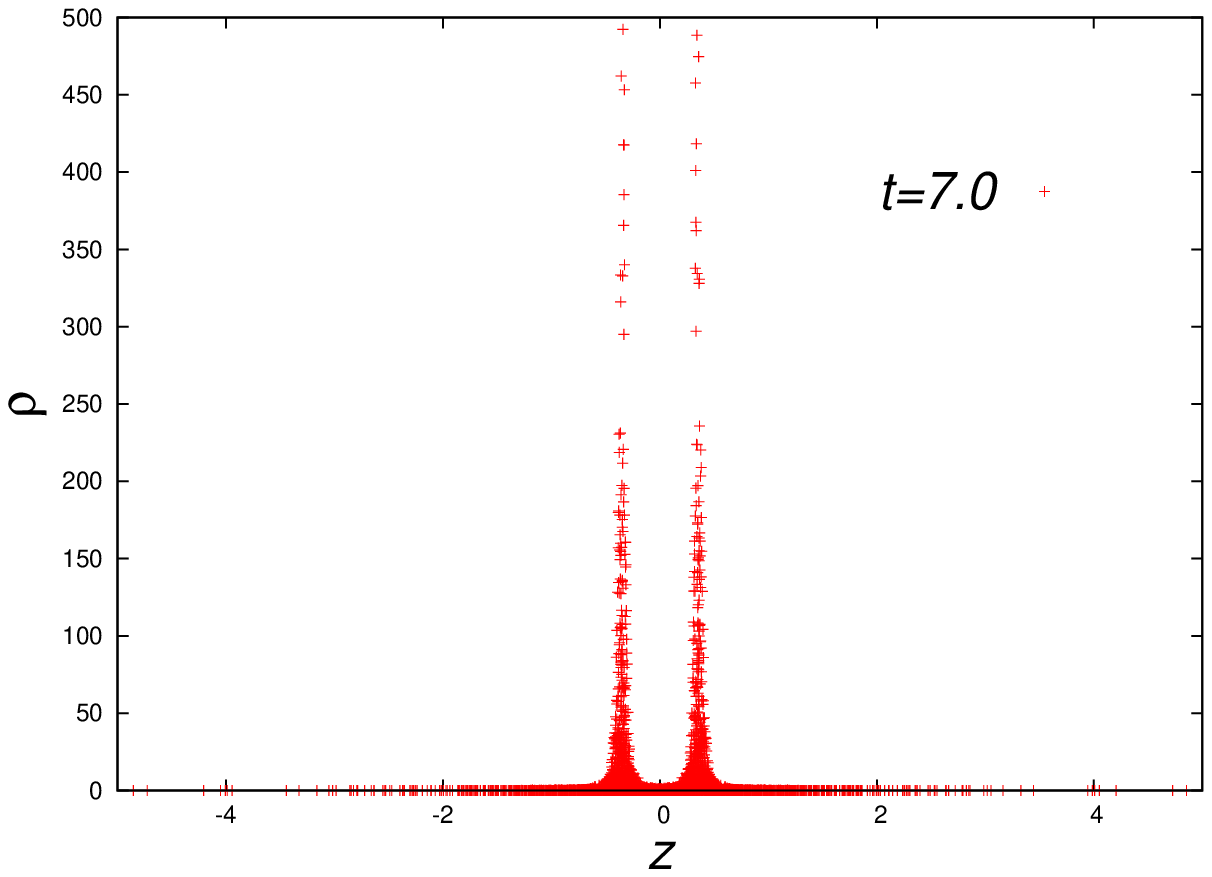}
\includegraphics[width=4cm]{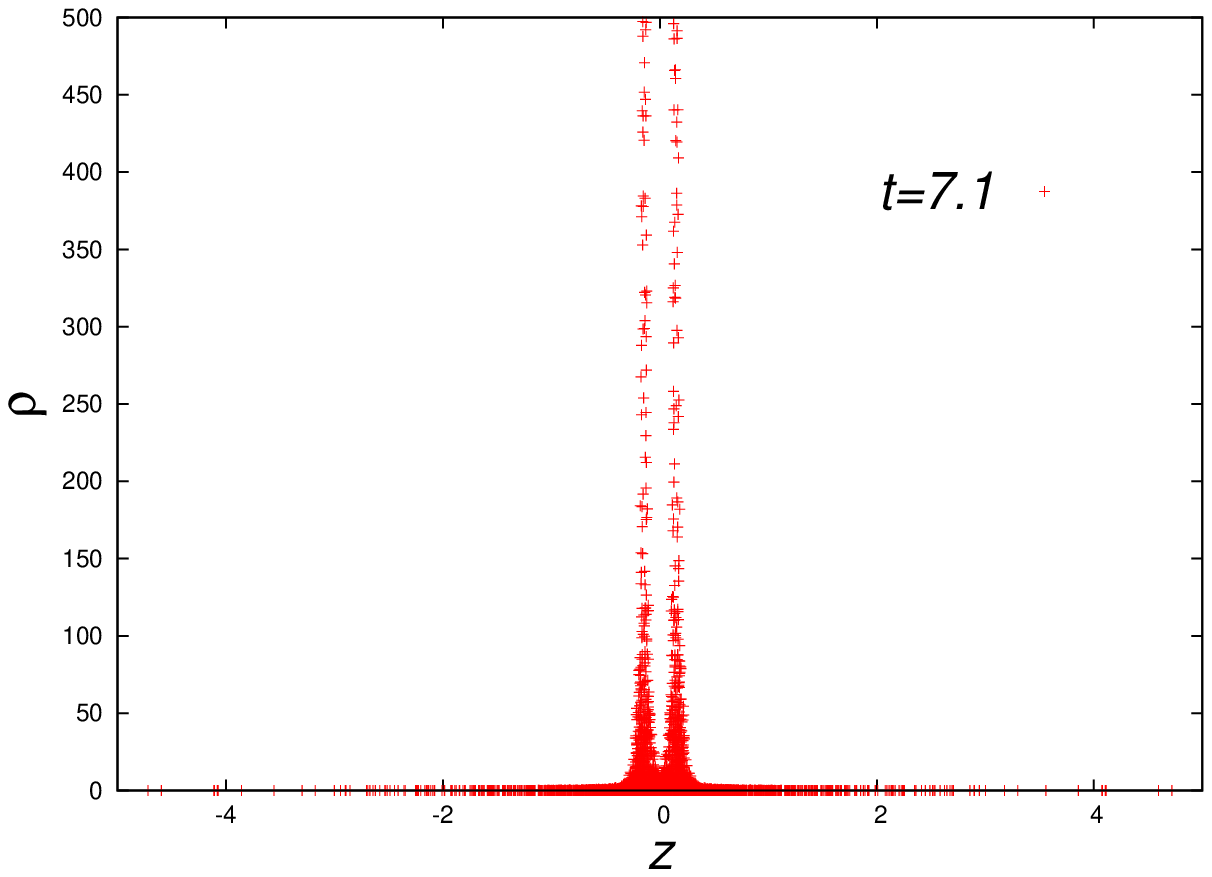}
\includegraphics[width=4cm]{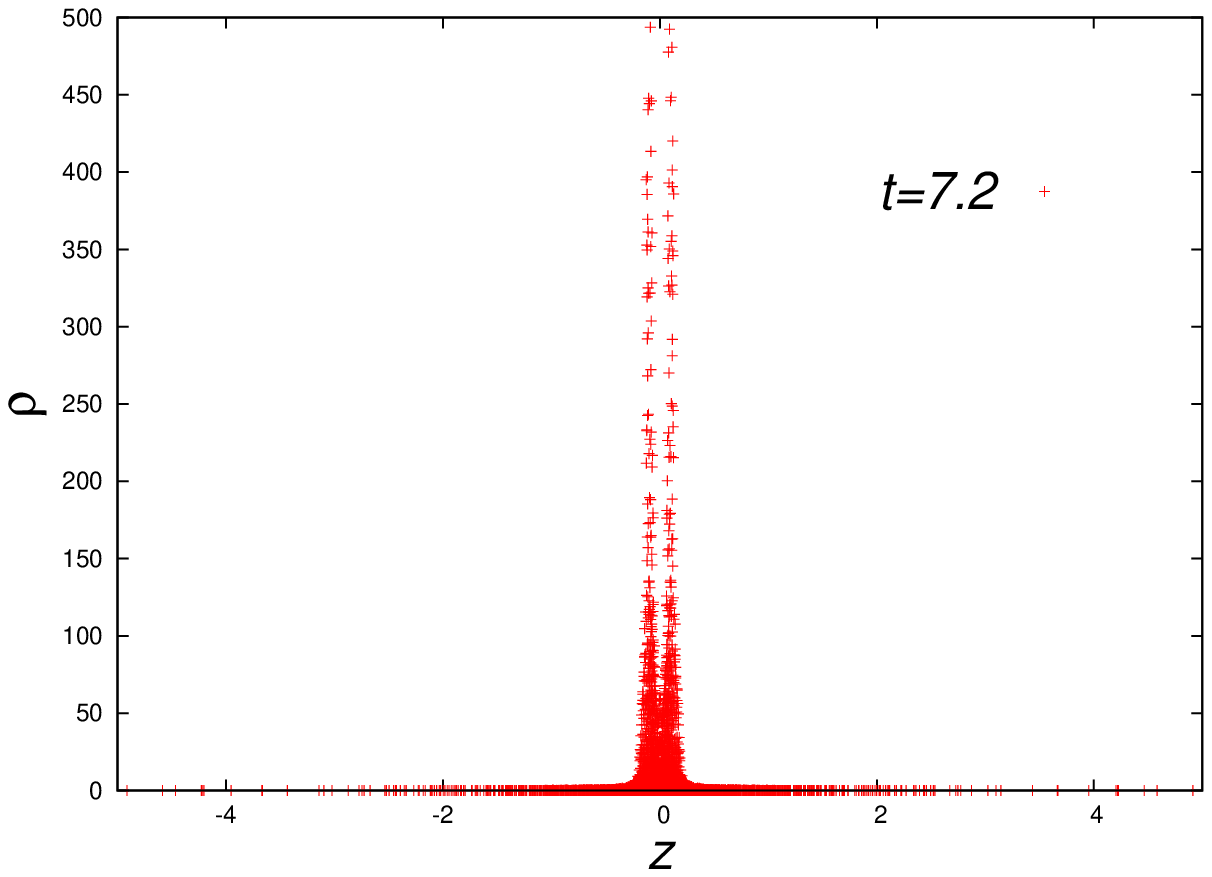}
\includegraphics[width=4cm]{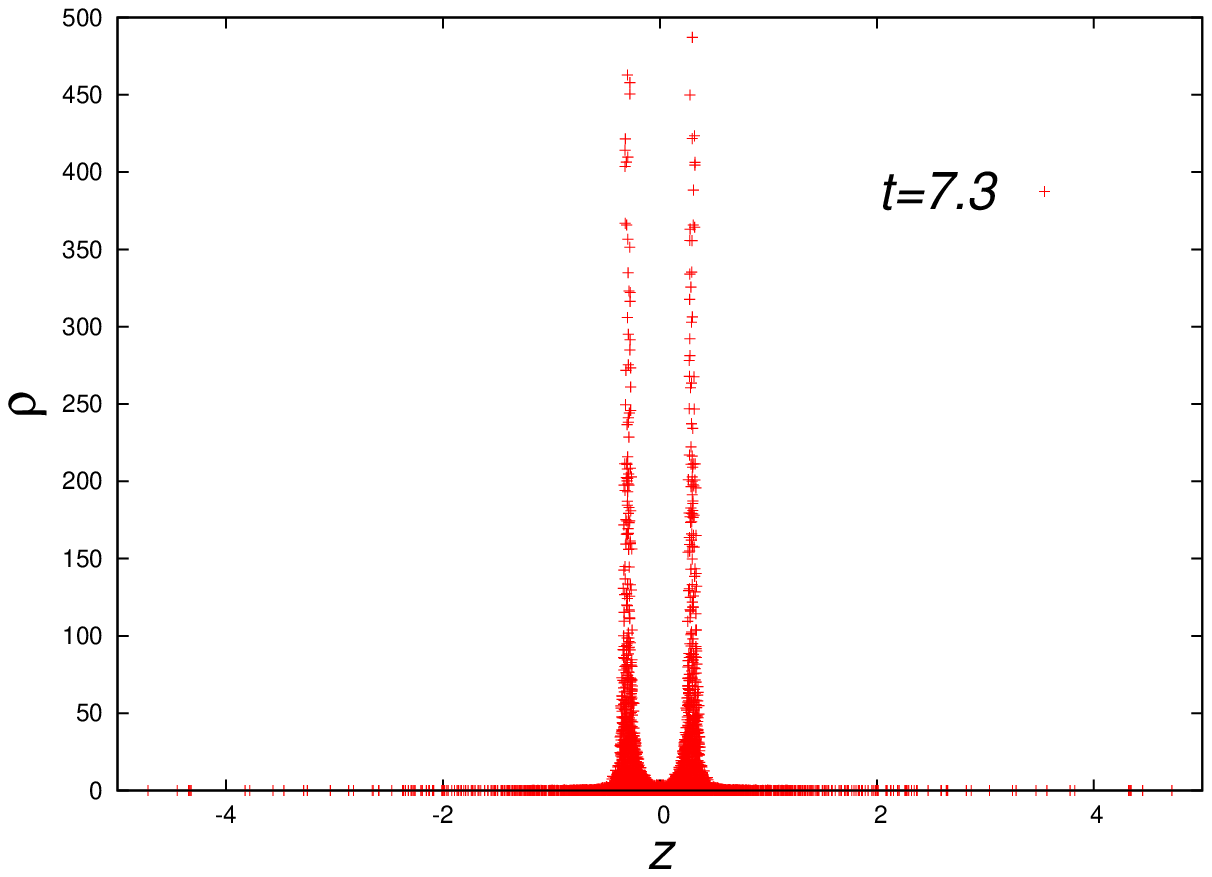}
\includegraphics[width=4cm]{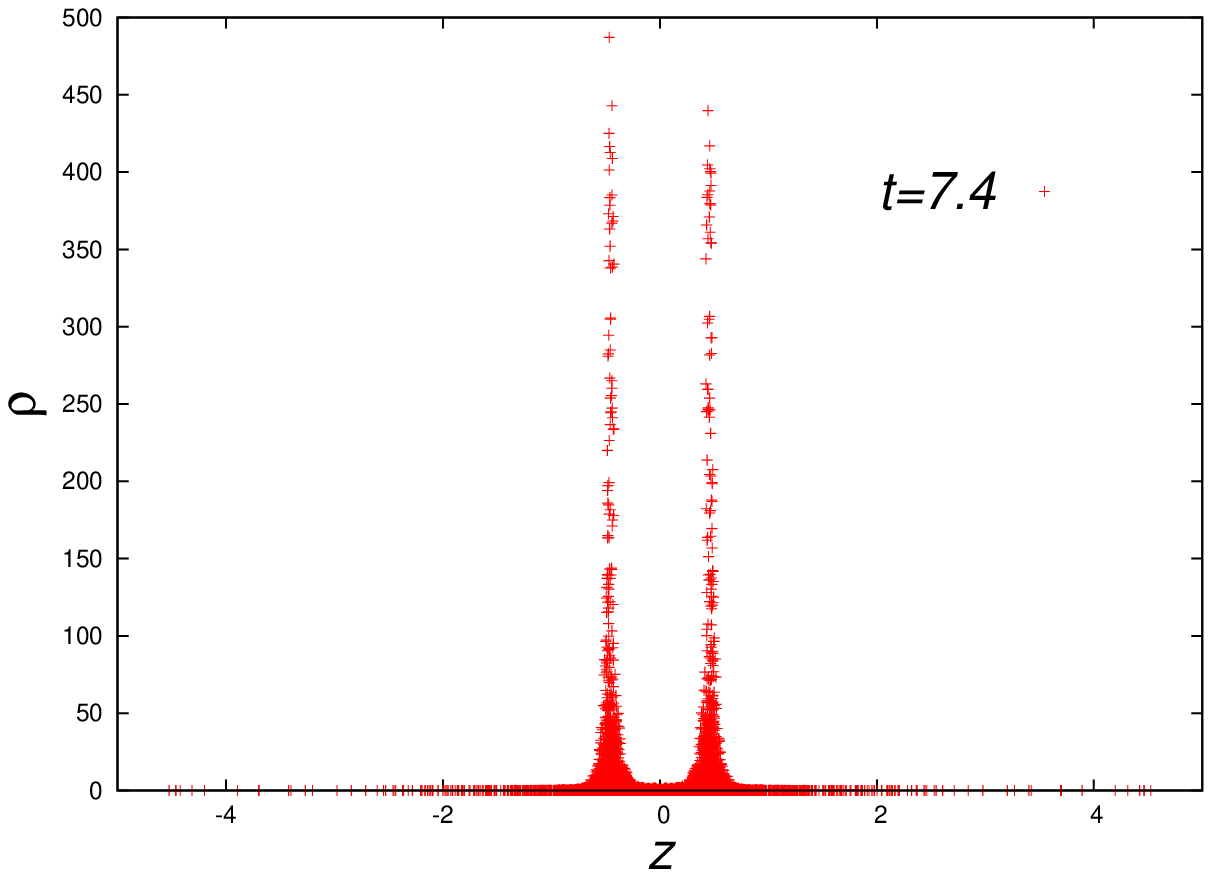}
\caption{\label{density_dust}In this plot we show snapshots at 
different times of the density as function of the $z$ coordinate of the 
structures of an ideal gas with zero pressure. The collision takes 
place after $t\sim7.2$. The evolution shows that the fluids cross each 
other without showing any pattern of interference.}
\end{figure}

\begin{figure}[htp]
\includegraphics[width=8cm]{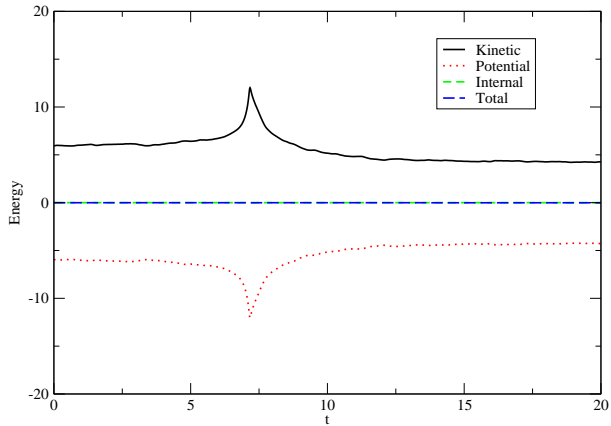}
\caption{\label{energy_dust}In this plot we show the total, kinetic, 
potential and thermal energy for the head-on collision of the 
pressureless fluid configurations. The peak appears approximately 
by the time the density of the fluid reaches its maximum.}
\end{figure}

The first case we analyzed is the pressureless fluid. In Fig. 
\ref{density_dust} we show snapshots at different times of the density 
as function of the $z$-position of the particles. The time of collision 
is around $t\sim7.2$ and it is clear that no interference pattern is 
generated during such a collision. In Fig. \ref{energy_dust} we show 
the kinetic, potential, internal and total energy of the system, in 
which it can be noticed that no internal energy is being involved in 
the model.

\begin{figure}[htp]
\includegraphics[width=4cm]{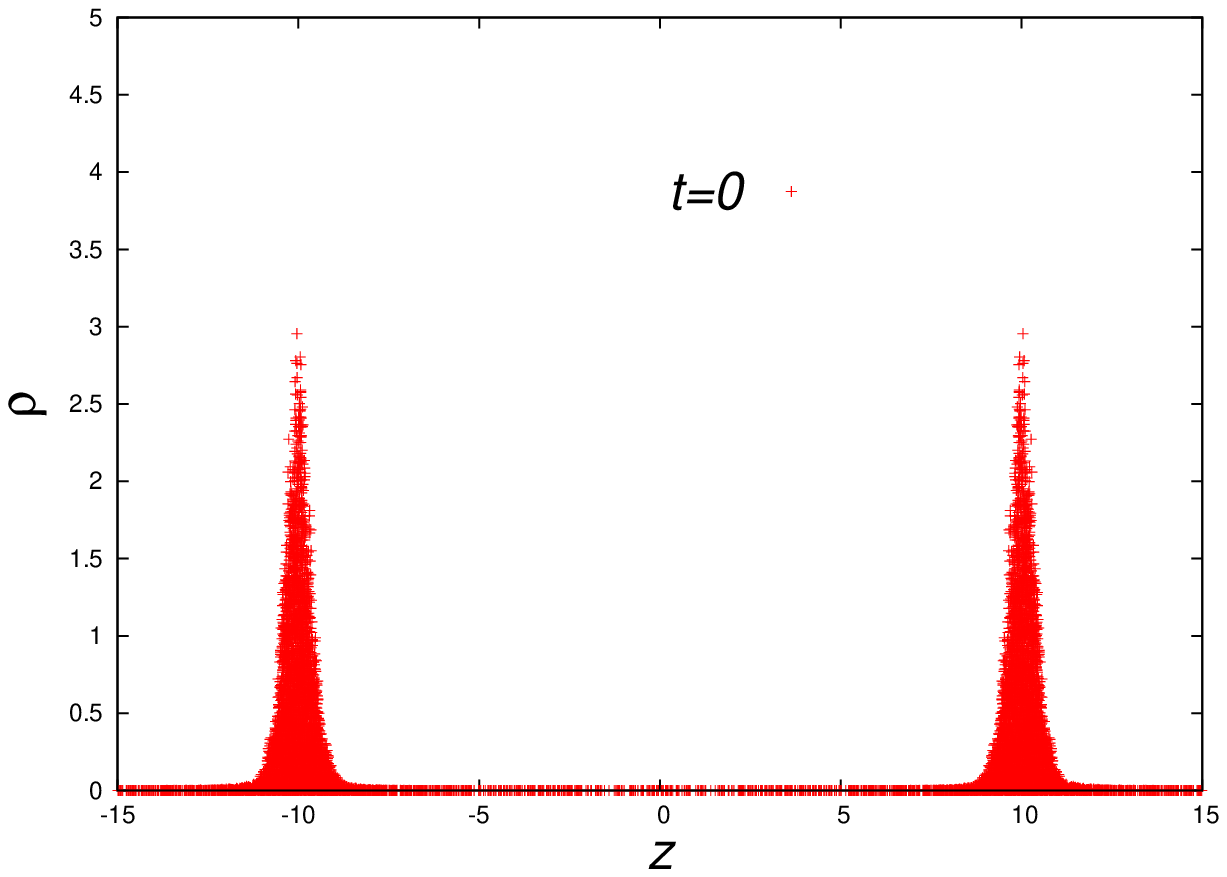}
\includegraphics[width=4cm]{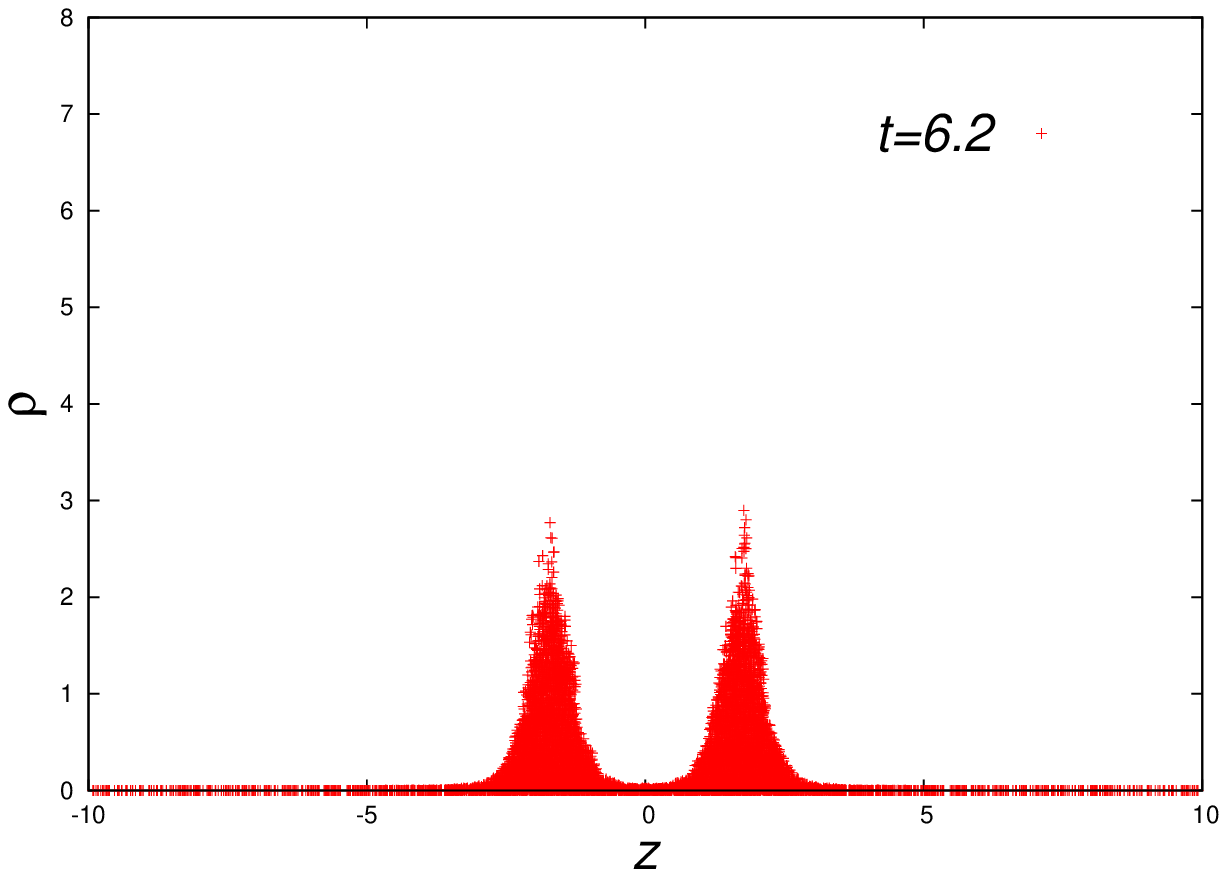}
\includegraphics[width=4cm]{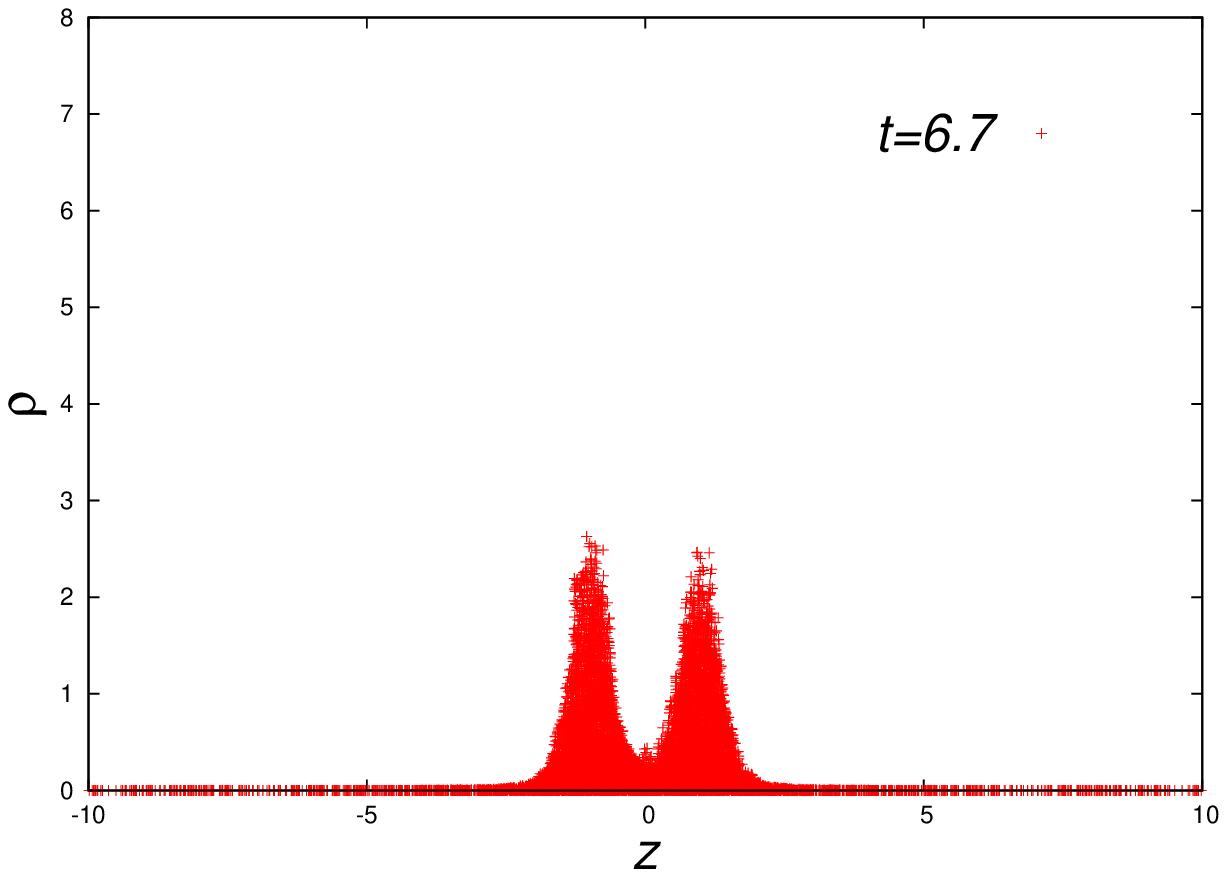}
\includegraphics[width=4cm]{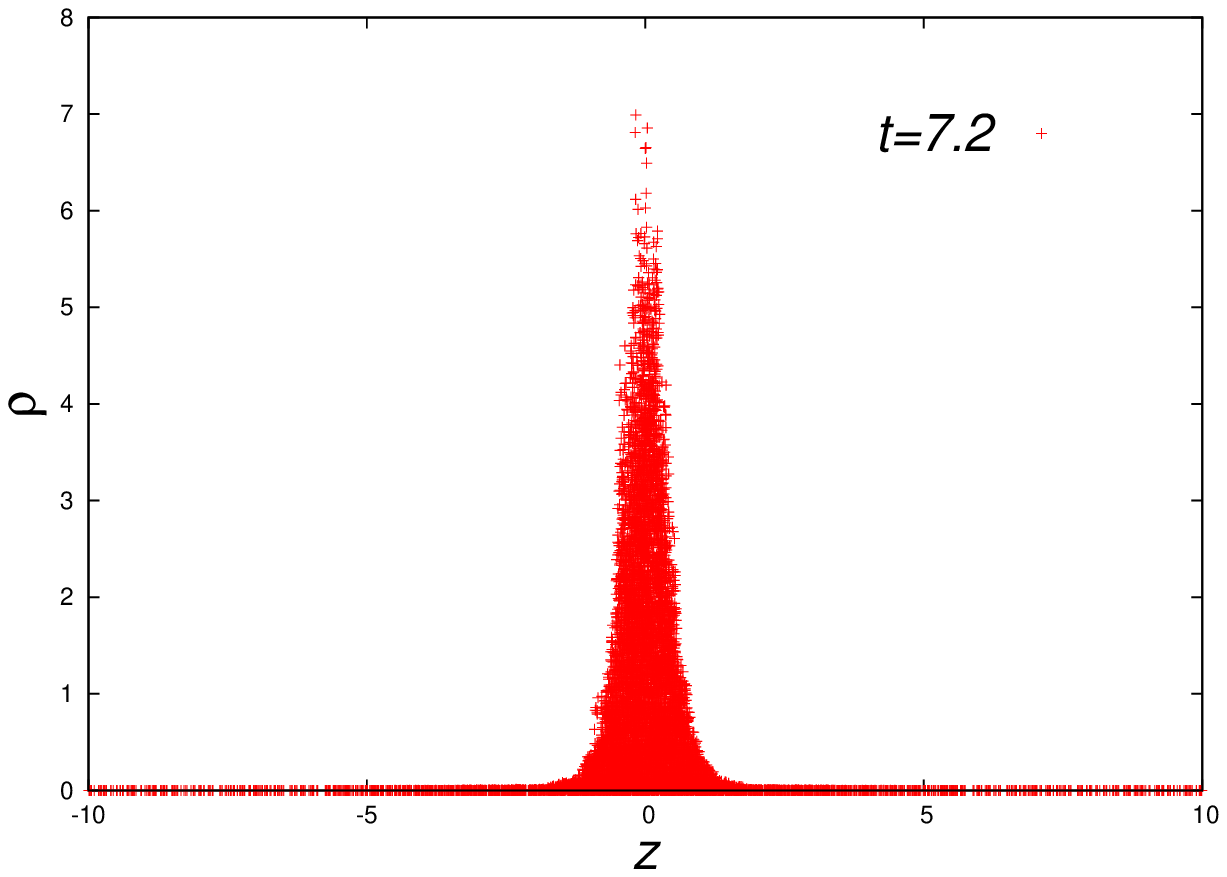}
\includegraphics[width=4cm]{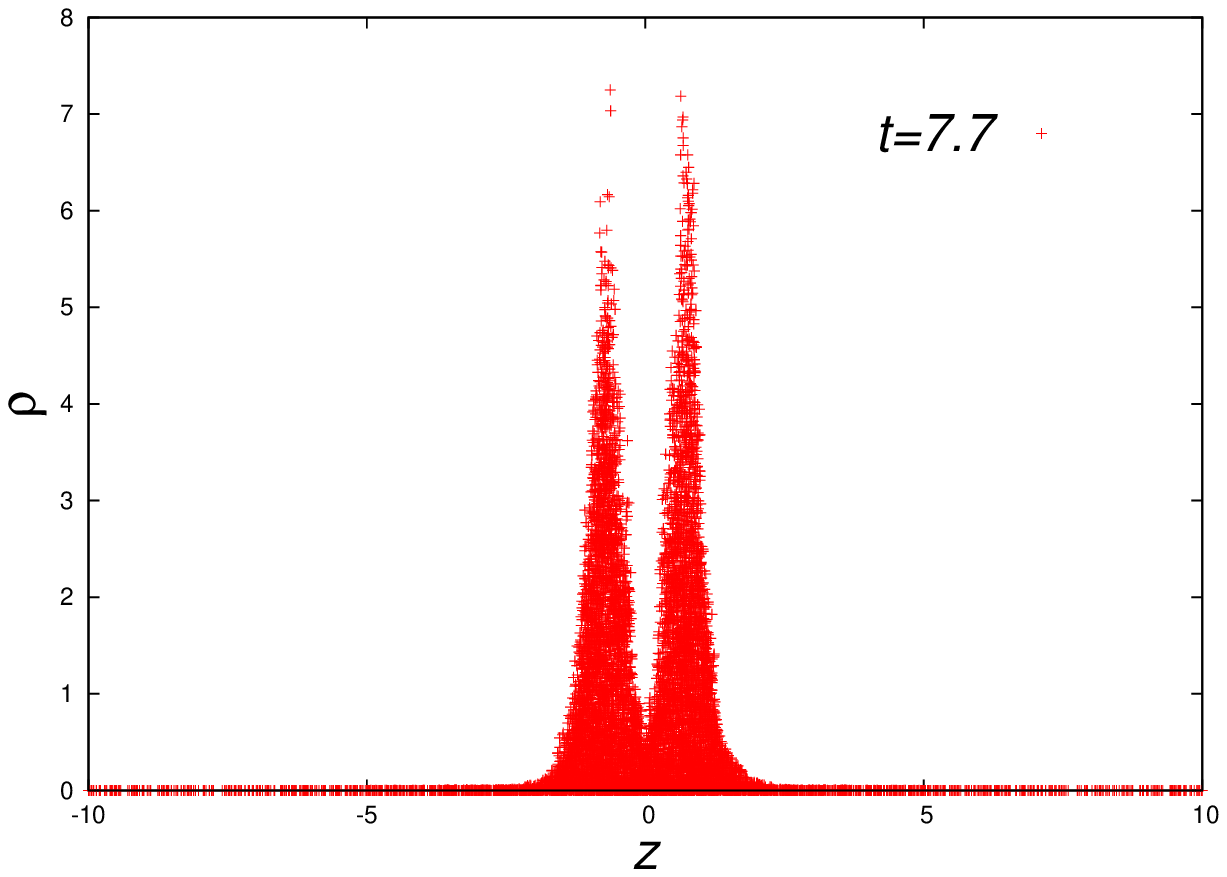}
\includegraphics[width=4cm]{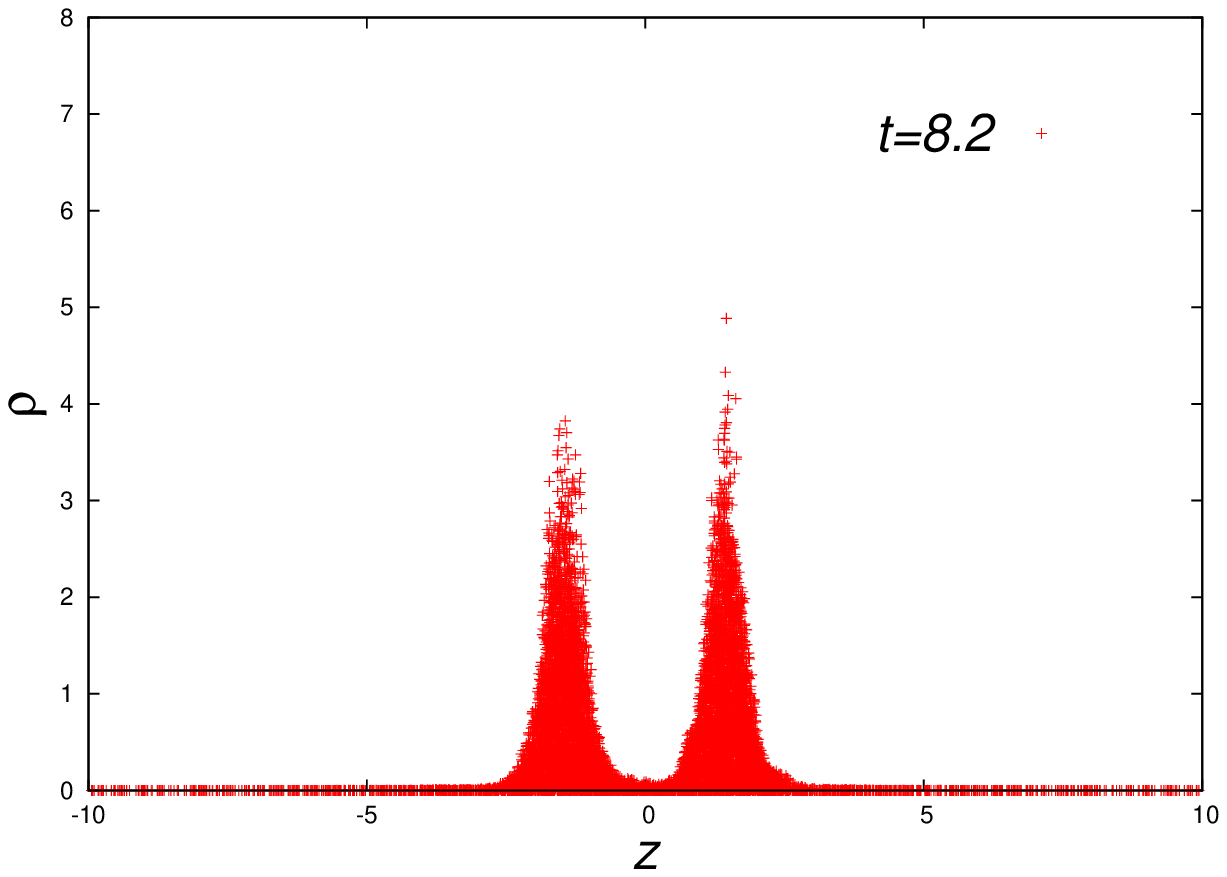}
\caption{\label{density_pol} In this plot we show snapshots at 
different times of the density as function of the $z$ coordinate of 
the structures of fluid. As in the pressureless case, the evolution 
shows the fluids crossing each other without showing any pattern of 
interference.}
\end{figure}

\begin{figure}[htp]
\includegraphics[width=8cm]{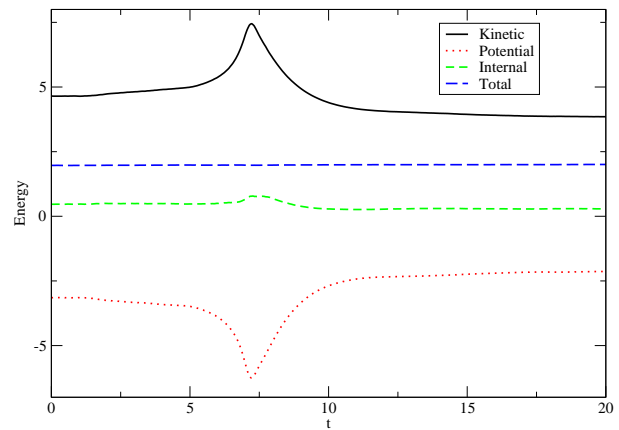}
\caption{\label{energy_pol}In this plot we show the total, kinetic, 
potential and thermal energy for the head-on collision of fluid with 
pressure.}
\end{figure}

The second case is the ideal gas case with pressure. As before, in 
Fig. \ref{density_pol} we present the density as function of the position 
for different times and in Fig. \ref{energy_pol} the energies of the 
system. In this scenario, again, we were unable to track any interference 
patterns. These figures illustrate the role played by the pressure, which 
produces the density profiles to be less steep compared to the 
pressureless case.

We have also performed simulations with various values of the initial 
linear momentum in the head-on direction and thus various values of the 
total energy of the system. We have also carried out our simulations 
with various numbers of particles and in all the cases the results are 
qualitatively the same.

\section{Conclusions}

We have solved the SP system as the equations describing the evolution 
of gravitating Bose condensates, which represents the BEC dark matter 
model and its properties at local scales. We have focused on the 
head-on collision of two equilibrium ground state configurations, 
and studied the interference pattern formation of the density of the 
configuration during the collision.

In order to investigate whether or not other type of fluids may show 
a similar interference pattern, we also studied the collision of two 
spherical configurations made of an ideal gas fluid, both with and 
without pressure using SPH techniques for different values of the 
total energy.

We found that the pattern formation during the collision of structures 
does not happen for the fluid. This leads to the conclusion that a 
fingerprint of the BEC dark matter model is the presence of 
interference patterns during the collision of structures. That is, 
if evidence of such interference patterns is not found, the BEC dark 
matter model should be ruled out.


\section*{Acknowledgments}

This work is supported by Grant Nos. 
CIC-UMSNH-4.9 and 4.23,
PROMEP UMICH-CA-22, UMICH-PTC-210,
CONACyT 79601 and 106466.
Our simulations were carried out in the IFM Cluster.




\end{document}